\documentstyle[twocolumn,prl,aps,epsfig,amssymb,floats]{revtex}
\def\beq{\begin{equation}}
\def\eeq{\end{equation}}
\def\bea{\begin{array}}
\def\eea{\end{array}}

\def\del{\partial }
\def\to{\rightarrow}

\def\[{\left[}
\def\]{\right]}
\def\({\left(}
\def\){\right)}

\def\sm0{{\widetilde{m}_0}}

\def\U1em{{U(1)_{\rm em}}}
\def\to{\rightarrow}

\def\sq2{\sqrt{2}}

\def\End{\end{document}}
\newcommand{\lae}{\stackrel{<}{\sim}}
\newcommand{\gae}{\stackrel{>}{\sim}}

\def\Journal#1#2#3#4{{#1} {\bf #2} (#4) #3}

\def\NPB{{\em Nucl. Phys.} B}
\def\PLB{{\em Phys. Lett.}  B}
\def\PRL{\em Phys. Rev. Lett.}
\def\PRD{{\em Phys. Rev.} D}
\def\ZPC{{\em Z. Phys.} C}
\def\EPC{{\em Euro. Phys. J.} C}

\begin{document}
\def\doublespaced{\baselineskip=\normalbaselineskip\multiply\baselineskip
  by 150\divide\baselineskip by 100}

\twocolumn[\hsize\textwidth\columnwidth\hsize\csname
@twocolumnfalse\endcsname

\title{
Testing Supersymmetry in the Associated Production of \\
CP-odd and Charged Higgs Bosons}
\author{
{\sc Shinya Kanemura}  \, {\rm and} \,
{\sc C.--P. Yuan}
}
\address{
\vspace*{2mm}
Department of Physics and Astronomy,
Michigan State University, East Lansing, Michigan 48824-1116, USA
}
\maketitle
\begin{abstract}

In the Minimal Supersymmetric Standard Model (MSSM),
the masses of the charged Higgs boson ($H^\pm$) and the
CP-odd scalar ($A$) are related
 by $M_{H^+}^2=M_A^2+m_W^2$. Furthermore,
because the coupling of
$W^-$-$A$-$H^+$ is fixed by gauge interaction, the
tree level production rate of
$q \bar q' \to W^{\pm \ast} \to A H^\pm$
depends only on one supersymmetry parameter --
the mass ($M_A^{}$) of $A$.
We show that to a good approximation this conclusion also
holds at the one-loop level. Consequently, this
process can be used to distinguish MSSM from its
alternatives (such as a general two-Higgs-doublet model)
by verifying the above mass relation,
and to test the prediction of the MSSM on the product
of the decay branching ratios of $A$ and $H^\pm$ in terms
of only one single parameter -- $M_A^{}$.

{PACS numbers:\,12.60.-i,\,12.15.-y,\,11.15.Ex
\hfill   ~~ [ \today ~and~ hep-ph/0112165 ]
}
\end{abstract}
]

\vspace*{2mm}

\setcounter{footnote}{0}
\renewcommand{\thefootnote}{\arabic{footnote}}

One of the
commonly discussed new physics models is 
the Minimal Supersymmetric Standard Model (MSSM). 
To describe an experimental
 data in the framework of the MSSM usually requires
introducing more than one
 supersymmetry (SUSY) parameters.
Hence, the usual practice is to compare data to
a pre-selected class of MSSM in which certain
well-defined relations among the SUSY parameters
 are assumed in order to reduce the number of independent
 variables needed for discussion.

An interesting question to ask is
``Can one find a process to test
SUSY models at colliders without making many assumptions on the
choice of SUSY parameters?''
To answer that,
let us consider the Higgs sector of the model.
In the MSSM, because of the supersymmetry,
two Higgs doublets have to be introduced in
 its Higgs sector.
 Although the MSSM Higgs sector
resembles the one in a type-II two-Higgs-doublet model (THDM)
\cite{hhg}, it has a very specific feature -- all the Higgs self
couplings $\lambda_i$ are fixed by the electroweak gauge couplings
$g$ and $g'$, as required by supersymmetry. Hence, at the tree
level, only two additional free parameters appear in the Higgs
sector of the MSSM. We may take $M_A$ (the mass of the CP-odd
Higgs boson $A$) and $\tan\beta$ (the ratio of the two vacuum
expectation values) as these two free parameters.

One of the striking features resulted from the requirement of supersymmetry
is that the mass ($m_h$) of the lightest CP-even Higgs boson ($h$)
has to be less than the mass ($m_Z$)
of the weak gauge boson $Z$ at the Born level,
although a large radiative correction due to a heavy top
quark can push this bound up to about 130\,GeV
in the MSSM \cite{h0mass}.
This result is interesting when compared
to the theoretical bounds on the mass of the SM Higgs boson.
Requiring the SM be a well defined theory
up to the Planck scale (about $10^{19}$\,GeV),
the Higgs boson mass has to be approximately between
130 GeV and 180 GeV \cite{smh0,quiros}.
Therefore, a light Higgs boson with its mass less than about 130 GeV
can be a signal of the supersymmetric models, especially the MSSM.
However, it is also known that such a light Higgs boson can
 exist in various non-SUSY models, such as
a general THDM or the Zee model, even when the cutoff scale of the
model is close to the Planck scale \cite{2hdm}. Hence, the
existence of a light Higgs boson by itself cannot rule out models
other than the MSSM (or its extensions).

Another striking feature resulted from the requirement of supersymmetry
is that the masses of the charged Higgs boson $H^\pm$
and the CP-odd scalar $A$ are strongly correlated.
At the Born level, they are related by the mass of the
$W^\pm$ boson ($m_W$) as

\vspace*{-2mm}
\noindent
\begin{eqnarray}
M^2_{H^\pm} = M^2_A + m^2_W.
\label{eq:massrel}
\end{eqnarray}

\vspace*{-2mm}
\noindent
For comparison, the corresponding mass relation in a general
THDM is $M^2_{H^\pm} = M^2_A + {1 \over 2} (\lambda_5 - \lambda_4) v^2 $,
where $v$ is the weak scale (246\,GeV) and
$\lambda_{4,5}$ are two free parameters of the model\cite{masses2}.
Therefore, the mass relation~(\ref{eq:massrel}) can be a
strong criterion to discriminate the MSSM from its alternatives,
e.g., a general THDM.

To test the mass relation~(\ref{eq:massrel}),
we propose to study the associated production of $A$ and $H^\pm$
at high energy hadron colliders, e.g.,
$p {\bar p} \to A H^\pm$ at the Fermilab Tevatron
(a 2\,TeV proton-antiproton collider) and
$p p \to A H^\pm$ at the CERN LHC (a 14\,TeV proton-proton
collider).
As to be explained below, this process has the following unique features:
({\it i}) its Born level rate generally depends on the masses of
    $A$ and $H^\pm$.
    Because of the mass relation (\ref{eq:massrel}), the
    MSSM prediction of the Born level rate only depends on one
    (in contrast to two or more) SUSY parameter -- $M_A$;
({\it ii}) the kinematic acceptance
    (therefore, the detection efficiency)
    of the signal events do not depend on the choice of other SUSY
    parameters because
    both $A$ and $H^\pm$ are spin-0 (pseudo-)scalar particles
    so that the kinematic distributions of their decay particles can be
    accurately modeled;
({\it iii}) it can constrain MSSM parameters by
    examining the product
    of the Higgs boson decay branching ratios (in contrast to the product of
     decay branching ratios {\it and} production rate);
({\it iv}) both $M_A^{}$ and $M_{H^\pm}^{}$
    can be reconstructed from its final
    state to test the mass relation~(\ref{eq:massrel});
({\it v}) finally, the electroweak radiative corrections to its
   production rate and to the mass relation ~(\ref{eq:massrel})
   are generally smaller than the expected experimental errors,
   such as the di-jet invariant mass resolution,

Either in the MSSM or the THDM, the coupling of
$W^\mp$-$A$-$H^\pm$\ is induced from the gauge invariant
kinetic term of the Higgs sector \cite{hhg}:

\vspace*{-2mm}
\noindent
\begin{eqnarray}
{\cal L}_{int}= \frac{g}{2} W^+_\mu (A \del^\mu H^- - H^- \del^\mu A)
+ {\rm h.c.} \, ,
\end{eqnarray}

\vspace*{-2mm}
\noindent
so that the coupling strength of $W^+H^-A$ is completely determined by the
weak gauge coupling $g$.
(In contrast, the coupling constants relevant to the interactions
of $W$-boson (or $Z$-boson) and
neutral Higgs bosons depend on $\beta$ and $\alpha$, where $\alpha$
characterizes the mixing between the two CP-even Higgs bosons
$h$ and $H$.)
Thus, the Born level
production rate of a $AH^\pm$ pair at hadron colliders
only depends on $M_A^{}$ and $M_{H^\pm}^{}$.
Since in the MSSM these masses are strongly correlated, cf.
Eq.~(\ref{eq:massrel}),
 the production rate of $p {\bar p},p p \to A H^\pm$
only depends on one SUSY parameter, which can be taken as $M_A$.

At the Tevatron and the LHC, the dominant constituent process
for the production of a $A H^\pm$ pair is
$q {\bar q'} \to W^{\pm\ast} \to A H^\pm$.
For a given $M_A$, the cross section
$\sigma(p {\bar p}, \, pp \to A H^\pm)$ is completely determined.
Its squared amplitude, after averaging over the spins and colors is
\begin{eqnarray}
{\overline {|{\cal M}|^2} } = {4 \over 3} m_W^4 G_F^2
{ {s} \over ({s}^2 -m_W^2)^2 + m_W^2 \Gamma_W^2}
P^2 \sin^2\theta \, , 
\end{eqnarray}
where $P=\sqrt{E_A^2-M_A^2}\, $ with
$E_A=({s}+M_A^2-M_{H^+}^2)/(2 \sqrt{{s}}) \, $
and
$\theta$ is the polar angle of $A$ in the center-of-mass (c.m.) frame
of $A$ and $H^\pm$.
We note that for the $c \bar b \to A H^+$ subprocess,
 in addition to the CKM (Cabbibo-Kobayashi-Maskawa)
suppressed $s$-channel $W$-boson diagram, there is
a $t$-channel diagram, that depends on $\tan \beta$.
However, the $c \bar b \to A H^+$ contribution
to the inclusive $AH^+$ rate is small. For example, its contribution
to the total rate is less than 0.01\% and 0.1\% at the
Tevatron and the LHC, respectively,
for $\tan \beta=40$ and $M_A=90$\,GeV.
For a smaller $\tan\beta$, its contribution becomes negligible.
Hence, we shall ignore its contribution in the following discussion.
In Fig.~1, we show the inclusive production
rate of $A H^\pm$ as a function of
$M_A$. Here, the CTEQ5M1 parton distribution
functions (PDF) \cite{cteq5m} are used and both the renormalization
and the factorization scales are chosen to be the invariant mass ($\sqrt{s}$)
of the $A H^\pm$ pair.
A next-to-leading order (NLO) QCD correction is also included,\footnote{
This is similar to the NLO QCD correction to the $W$-boson
production at hadron colliders, except at a different invariant mass.
}
which typically increases the LO rate by about 20\% (when the
same set of PDF is used).

It is trivial to model the kinematic acceptance (therefore, the
detection efficiency) of the signal event.
This is because both $A$ and $H^\pm$ are spin-0 bosons.
 Therefore, if the signal is not found,
 knowing the luminosity of the collider,
 the detection efficiency, and the theoretical production rate,
one can conclude from the data a
constraint on the product of the decay branching ratio of $A$ and
$H^\pm$ as a function of $M_A$.
 For example,if the decay mode of
$A \to b \bar b$ and $H^+ \to \tau^+ \nu_\tau$
 is studied and no excess is found for a given mass bin of
$M_A$ (hence, $M_{H^\pm}$) when
comparing with the experimental data, then one can constrain
the MSSM by demanding the product of
the branching ratios, ${\rm Br}(A \to b {\bar b}) \times {\rm
Br}(H^+ \to \tau^+ \nu_\tau)$, to be bounded from above as a
function of $M_A$. Needless to say that applying the same strategy, one
can constrain the product ${\rm Br}(A \to X) \times {\rm Br}(H^+ \to
Y)$ for any decay mode $X$ and $Y$ predicted by the MSSM
as a function of only one SUSY parameter -- $M_A$.

In case that a signal is found, the analysis is slightly more
complicated.
In the MSSM, the mass of the heavier CP-even Higgs boson ($H$) is
not very different (less than about 10\,GeV)
from $M_A$ for $M_A \gae 120$ GeV and $\tan\beta \gae
10$. In this case, $q{\bar q'} \to H H^\pm$ can produce the similar final
states as $q{\bar q'} \to A H^\pm$.
Generally, the coupling of $W^\pm H H^\mp$ depends on $g$ and
$\sin(\alpha-\beta)$.
However, for $M_A \gae 190$ GeV and $\tan\beta \gae 10$,
$\sin^2(\alpha-\beta)\simeq 1$ and
the production rate of $HH^\pm$ is almost the same as $AH^\pm$
in the MSSM.
When both of them decay into the same decay channels,
it will be very difficult to separate the production of $AH^\pm$ from
$H H^\pm$ unless a fine mass resolution can be achieved experimentally.
Nevertheless, studying different decay channels can help to
separate these two production modes. For instance, a heavy
$H$ can decay into
a $ZZ$ pair at the Born level, but $A$ cannot.

In conclusion, if no signal is found experimentally,
a {\it conservative} bound on the product
of the decay branching ratios of $A$ and $H^\pm$ can be derived
for a CP-conserving model.
This is because in a CP-conserving model,
the $AH^+$ and $HH^+$ production modes do not interfere even if the masses
of $A$ and $H$ are about the same.
(We note that $A$ is a CP-odd scalar, while $H$ is CP-even.)

To test the MSSM relation~(\ref{eq:massrel})
via the $AH^\pm$ production process is straightforward.
For example, let us consider the $b {\bar b} \tau \nu$
mode.
If the signal is sufficiently large as compared to the backgrounds,
a resonance bump should be observed (at a value
$\langle M_{b\bar{b}} \rangle$) in the distribution of the
$b\bar b$ invariant mass.
Then, by searching for the corresponding Jacobian peak
(at a value $\langle M_{T}(\tau\nu) \rangle$) in the
distribution of the transverse mass of the
$\tau\nu$ pair, one can test the MSSM
by examining whether
$\langle M_{T}(\tau\nu) \rangle$
is consistent with $\sqrt{\langle M_{b\bar{b}} \rangle^2 + m_W^2}$
within the accuracy of the mass resolution of the
detector.
By testing this mass relation via the process
$p\bar p,pp \to A H^\pm \,$, one can
discriminate the MSSM from its alternative (e.g., a general THDM).
\begin{figure}[t]
\label{fig:cross}
\begin{center}
\epsfig{file=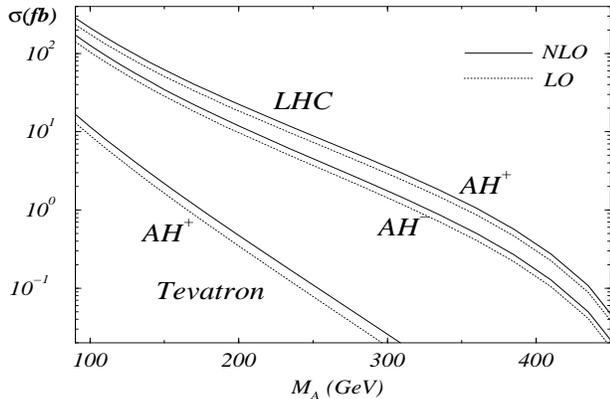,width=8.0cm,height=5.5cm}
\end{center}
\vspace*{-5mm}
\caption{The LO (dotted lines) and NLO QCD (solid lines)
cross sections of the $AH^+$ and $AH^-$ pairs
as a function of $M_A^{}$.
The cross sections for $AH^+$
and $AH^-$ coincide at the
Tevatron for being a $p \bar p$ collider. }
\end{figure}

In order to prove that the proposed process can be used to test
the MSSM and is sensitive to only one SUSY parameter -- $M_A$, we
have to show that the SUSY electroweak correction, which occurs at
the one-loop level, to the production rate is small. (The final
state SUSY QCD correction does not contribute until the two-loop
level.) Specifically, we need to consider two essential points:
($a$) what is the typical size of the radiative correction to the
coupling of $W^\mp$-$A$-$H^\pm$? ($b$) does the mass
relation~(\ref{eq:massrel}) hold beyond the Born level? In the
following, we shall have a more detailed discussion on these
important questions. A brief summary is that as long as the
typical SUSY mass scale is at the order of a few TeV or below, the
one-loop correction to the $W^\pm A H^\mp$ vertex can modify the
production rate of the $A H^\pm$ pair at most by a few percent.
Therefore, the electroweak correction is smaller than (i) the
expected statistical and systematic errors of the experimental
data, or (ii) the uncertainty in the theory prediction of the
production rate originated from the parton distribution functions
(which is estimated to be about 6\% at the Tevatron and 5\% at the
LHC for $M_A^{}=120$ GeV, when applying the prescription presented
in Ref.~\cite{pumplin}), or (iii) the higher order ($\alpha^2_s$
or above) QCD correction (which is estimated to be about 10\% at
the Tevatron and less than a percent at the LHC for $M_{A}^{}=120$
GeV, when varying the factorization scale around $\sqrt{{s}}$, the
c.m. energy of the subprocess $q{\bar q'} \to A H^\pm$, by a
factor of 2). Furthermore, as long as the typical SUSY mass scales
are at the order of a couple of TeV or below, the radiative
correction to the mass relation~(\ref{eq:massrel}) is generally
smaller than the typical mass resolution of the experimental
measurement (which is about 10\,GeV for a $100$\,GeV Higgs boson
decaying into jets).

The dominant one-loop electroweak corrections to the
$q{\bar q'} \to A H^\pm$ process come from the loops of top ($t$) and
bottom ($b$) quarks as well as their supersymmetric partners, i.e.
stops ($\tilde{t}_{1,2}$) and sbottoms ($\tilde{b}_{1,2}$), in the
MSSM. This is due to their potentially large couplings to Higgs bosons.

In the following, we discuss the quark- and squark-loop
radiative corrections to the effective coupling of $W^\pm A H^\mp$ and
to the mass relation~(\ref{eq:massrel}).
In our calculation, we adopt the on-shell renormalization scheme
developed by Dabelstein in Ref.~\cite{dabelstein}
(see Appendix {\bf A}).

The part of one-loop effective coupling of
$W^\pm A H^\mp$, that is relevant to the production
process $q {\bar q'} \to H^+A$, can be written as\footnote{
The other form factor, $(p_{A}^{} + p_{H}^{})^\mu$,
does not contribute to this process for massless quarks.
}

\vspace*{-2mm}
\noindent
\begin{eqnarray}
  M_{WHA}^\mu(q^2) &=&
 -\frac{{\bar g}}{2} (p_{A}^{} - p_{H}^{})^\mu \,
  \left[ 1 + F^{(1)}(q^2) \right],
\end{eqnarray}

\vspace*{-2mm}
\noindent
where  $q^\mu$, $p_A^\mu$ and
$p_H^\mu$ are the incoming momenta of
$W^+$, $A$ and $H^-$, respectively,
and ${\bar g}$ is the effective weak gauge coupling evaluated
at $q^2$.
Hence, the radiative correction to the cross section of the sub-process
$q {\bar q'}\to AH^+$ at the one-loop order is
\begin{eqnarray}
  K^{(1)}(q^2)  \equiv 2 {\rm Re}\, F^{(1)}(q^2).
\label{eq:kfactor}
\end{eqnarray}

The detailed calculation for $F^{(1)}(q^2)$ is summarized in
Appendix~{\bf B}.
As shown in Eqs.~(\ref{eq:qform1}), (\ref{eq:qform2}) and (\ref{eq:qform3}),
the quark-loop contribution is proportional to
the squared Yukawa coupling constants
$y_t^2 (= 2 m_t^2 \cot^2\beta/v^2)$ and
$y_b^2 (= 2 m_b^2 \tan^2\beta/v^2)$.
In the large $m_t$ or large $m_b \tan\beta$ limit,
it can be written as

\vspace*{-2mm}
\noindent
\begin{eqnarray}
 F^{(1)}_{\rm quark}
 &\sim& \frac{N_c}{16\pi^2}
 \left[ -\frac{1}{4} y_t^2
  +  \frac{1}{2} \left( \frac{3}{2} - \ln \frac{m_t^2}{m_b^2} \right)
        y_b^2  \right],
\end{eqnarray}

\vspace*{-2mm}
\noindent
where $N_c(=3)$ is the number of colors.
Since $y_t^2$ and $y_b^2$ are at most ${\cal O}(1)$
for $\tan\beta\simeq 1$ and $m_t/m_b$, respectively,
$F^{(1)}_{\rm quark}$ is at most a few percent for
$1 \lae \tan\beta \lae m_t/m_b$.

We also calculate the squark-loop contribution.
As compared to the quark effects, the squark effects are rather complex
due to the additional
free (SUSY) parameters.
The mass eigenstates $\tilde{f}_{1,2}$
($\tilde{f}=\tilde{t}$ or $\tilde{b}$)
of the squarks
are obtained from the weak eigenstates
$\tilde{f}_{L,R}$ by diagonalizing the mass matrices
defined through \cite{hk}

\vspace*{-2mm}
\noindent
\begin{eqnarray}
  {\cal L}_{\rm mass} = - (\tilde{f}_L^\ast, \tilde{f}_R^\ast)
        \left( \begin{array}{cc}
               M_L^2 & m_{f} X_{f}\\
               m_{f} X_{f} & M_R^2\\
         \end{array} \right)
        \left( \begin{array}{c}
              \tilde{f}_L \\ \tilde{f}_R
         \end{array} \right),
\end{eqnarray}
\noindent
where, $M_L^2=M_{\tilde{Q}}^2+m_f^2+
       (m_Z^2 \cos 2\beta) (T_{f_L^{}}-Q_f  s_W^2)$ and
       $M_R^2=M_{\tilde{U}, \tilde{D}}^2+m_f^2+
       (m_Z^2 \cos 2\beta) Q_f s_W^2$.
In this expression,
$M_{\tilde{Q}}^2$, $M_{\tilde{U}}^2$ (for $\tilde{f}=\tilde{t}$)
and $M_{\tilde{D}}^2,$ (for $\tilde{f}=\tilde{b}$)
are the soft-breaking masses for
$\tilde{f}_L$, $\tilde{t}_R$ and $\tilde{b}_R$, respectively;
$s_W=\sin \theta_W$ with $\theta_W$ being the weak mixing angle;
$T_{f_L^{}}$ and $Q_f$ are the isospin and the electric charge
of the quark $f_L$.
Moreover,
$X_t =   A_t - \mu \cot\beta$ and
  $X_b =   A_b - \mu \tan\beta$,
where $A_t$ ($A_b$) is the trilinear $A$-term for $t$ ($b$), and
$\mu$ is the SUSY invariant higgsino mass\cite{hk}.
For completeness, we have listed all the relevant squark and Higgs bosons
couplings in {Appendix~{\bf C}}, so that
the squark-loop contributions to $F^{(1)}(s)$,
cf.~(\ref{eq:ff}), can be directly read
out from the
Eqs.~(\ref{eq:sqform1}), (\ref{eq:sqform2}) and (\ref{eq:sqform3}).

To examine the effect of one-loop electroweak corrections,
we shall discuss two limiting cases below.
Firstly, we consider the cases with $\mu=A_t=A_b=0$, i.e., the
cases without stop mixing ($|X_t|=0$) and sbottom mixing
($|X_b|=0$). Under this scenario, the masses of squarks are
proportional to $M^2$, and all the relevant couplings between
squarks and Higgs bosons are independent of the soft-breaking
masses $M_{\tilde{Q}}$, $M_{\tilde{U}}$ and $M_{\tilde{D}}$ (see
Appendix~{\bf C}). Thus, the squark-loop effect is decoupled and
its contribution is very small for a large value of $M$, where $M
\equiv M_{\tilde{Q}} \simeq M_{\tilde{U}} \simeq M_{\tilde{D}}$.
(Throughout this paper we denote $M$ as the typical scale of the
soft-breaking masses.) For a smaller $M$, $F^{(1)}_{\rm squark}$
becomes larger. However, $M$ cannot be too small because a small
$M$ implies light squarks whose masses are already bounded from
below by the direct search results\cite{PDG}. Furthermore, as to
be shown later, the case with a small $M$ is also strongly
constrained by the $\rho$ parameter measurement.
Secondly,
we examine the case with a large stop mixing, assuming
$m_t |X_t| \sim M^2 \gg m_Z^2$.
Such a large stop mixing leads to a large mass splitting between
$\tilde{t}_1$ and $\tilde{t}_2$ so that
$m _{\tilde{t}_1} \simeq {\cal O}(m_Z)$ and
$m_{\tilde{t}_2} \simeq \sqrt{2} M$, while
$m_{\tilde{b}_{1,2}} \simeq M$.
The leading squark contribution to $F^{(1)}(q^2)$
can be expressed as

\vspace*{-3mm}
\begin{eqnarray}
\!\!\!\!\!\!\!\!\!\!\!\!\!\!\!\!\!\!\!\!
\!\!\!\!\!\!\!\!\!\!\!\!\!\!\!\!\!\!\!\!
  F^{(1)}_{\rm squark}
 \sim \frac{-N_c}{16\pi^2}
 \left[  \left(\frac{3}{4}-\ln{2} \right)
    \left(\frac{Y_{\tilde t}}{M}\right)^2
\right.\nonumber
\end{eqnarray}

\vspace*{-8mm}
\begin{eqnarray}
\label{eq:fsquark}
  \;\;\;\;\;\;\;\;\;\;\;\;\;\;\;\;\;\;
  \;\;\;\;\;\;\;\;\;\;\;\;\;\;\;\;\;\;
 \left.
+ \left( \frac{13}{6} -3 \ln{2} \right)
      \left(\frac{Y_{\tilde b}}{M} \right)^2  \right],
\end{eqnarray}

\vspace*{-3mm}
\noindent
with
 $Y_{\tilde{t}} =  \frac{m_t}{v} (A_t \cot\beta+\mu)$ and
 $Y_{\tilde{b}} =  \frac{m_b}{v} (A_b \tan\beta+\mu)$.
Since in this case $|A_t| \simeq |M^2/m_t \pm |\mu| \cot\beta|$,
we have $|Y_{\tilde{t}}|
\lae {\cal O} (M^2/v)$
for $|\mu| \lae M$ and $1 \lae \tan\beta$.
When $|A_b| \simeq |A_t| > |\mu|$ and $\tan\beta \lae m_t/m_b$,
we find $|Y_{\tilde{b}}| \lae {\cal O} (M^2/v)$.
Thus, with a large stop mixing ($m_t |X_t| \simeq M^2$),
$F^{(1)}_{\rm squark}$ is proportional to the soft-breaking
mass scale $M$, and does not decouple in the large $M$ limit.
However, the $\rho$-parameter (or the $T$-parameter) can
also strongly
constrain such kind of model.
With a large stop mixing ($M^2 \simeq m_t |X_t|$),
the squark contribution (cf. Appendix~{\bf D})
to the $\rho$-parameter is

\vspace*{-2mm}
\begin{eqnarray}
 \Delta \rho_{\rm squark}
 \simeq
(2.2\times 10^{-3}) \frac{M^2}{v^2}.  \label{eq:rho}
\end{eqnarray}

\vspace*{-2mm}
\noindent
Since any new physics contribution to the $\rho$-parameter
has to be bounded by data as \cite{rho}

\vspace*{-2mm}
\noindent
\begin{eqnarray}
  -1.7 < \Delta  \rho_{\rm new} \times 10^3  < 2.7,
  \;\;\;\; {\rm at} \,2\sigma \,{\rm level,}  \label{rho}
\end{eqnarray}

\vspace*{-2mm}
\noindent
the scale $M$ cannot be too large in this case.
Consequently, the above $F^{(1)}_{\rm squark}$
is constrained to be
smaller than a few percent as long
as  $\mu^2$ is not much larger than $M^2$.

To examine the effect from the stop and sbottom
loops to the production rate of $AH^\pm$, we consider 4 sets of
SUSY parameters, as listed in Table~1,
which give the largest allowed deviation in the
$\rho$-parameter.
Set~1 and Set~2 represent the cases without either a stop
mixing ($X_t=0$) or a sbottom mixing ($X_b=0$), and
Set~3 and Set~4 are the cases with a large stop
mixing ($m_t |X_t| \simeq M^2$) and $m_{\tilde{t}_1}\simeq
100$ GeV.
The $K^{(1)}(s)$ factor,
as defined in Eq.~(\ref{eq:kfactor}),  is shown
in Fig.~2 as a function of the invariant mass ($\sqrt{s}$) of
the constituent process for $M_A^{}=90$ GeV.
It is clear that the quark-loop contribution to $K^{(1)}(s)$
dominates the squark-loop contribution.
For the above sets of SUSY parameters (Set 1-4),
the squark-loop contribution is smaller than
the quark-loop contribution
by about a factor of 100.
Generally, the squark contributions are at most a few percent,
unless $|\mu|$ is taken to be very large as compared to
the scale $M$.
We have checked that
this conclusion does not change when our
assumption of $M_Q^2 \simeq M_U^2 \simeq M_D^2$
is relaxed to some extent.
Including both the quark- and squark-loop contributions to $K^{(1)}(s)$,
we found that the correction
to the hadronic cross section of $H^+A$ production
in the invariant mass region just above the $H^+A$ threshold,
where the constituent cross section is the largest, is at
a percent level.
In summary, we illustrated that to be consistent
with the low-energy data
and the direct search results for stops and sbottoms,
the one-loop electroweak correction to the production rate
of $pp, \, p {\bar p} \to A H^\pm$ is small (at most a few percent).

\begin{table}
\begin{tabular}{|c|cccc|}
 & Set1 & Set2 & Set 3 & Set4  \\ \hline
$M_{\tilde{Q}}=M_{\tilde{U}}=M_{\tilde{D}}$ (GeV)
& 106 & 84 & 408 & 409 \\
$\tan\beta$ &2 &40 &2 &40 \\
$A_t=A_b$ (GeV) & 0 & 0 & $+1261$ & $+1119$ \\
$\mu$ & 0 & 0 & +300 & +300 \\ \hline
$m_{\tilde{t}_1}$ (GeV) &197 & 184 & 100 & 100\\
$m_{\tilde{t}_2}$ (GeV) &199 & 188 & 612 & 611\\
$m_{\tilde{b}_1}$ (GeV) &108 & 88  & 407 & 373\\
$m_{\tilde{b}_2}$ (GeV) &116 & 103 & 412 & 447 \\ \hline
$\Delta\rho_{\rm squarks} \times 10^3$   & 2.72 & 2.70 & 2.71 & 2.70 \\
\end{tabular}
\vspace*{2mm}
\caption{The SUSY (input and output) parameters used in Fig.~2.}
\end{table}

\begin{figure}[t]
\begin{center}
\epsfig{file=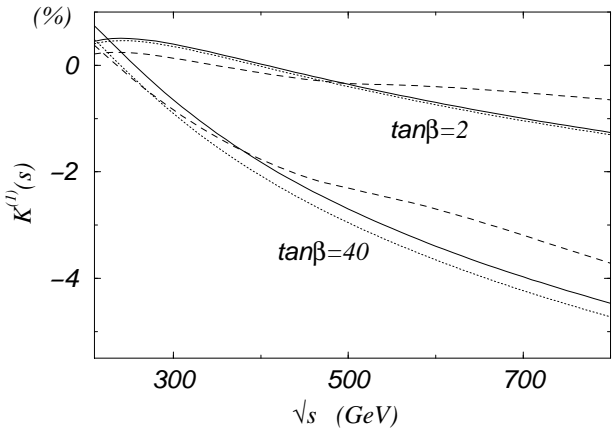,width=8cm,height=6cm}
\end{center}
\vspace*{-5mm}
\caption{The $K$-factor, $K^{(1)}(s)$, of the
constituent process $q {\bar q'} \to H^+A$ for $M_A^{}=90$ GeV,
as a function of the invariant mass $\sqrt{s}$ of $q {\bar q'}$.
The solid lines correspond to the top and bottom quark contribution.
The cases where the squark-loop contribution is included are
described by the dotted lines
for those without stop mixing (Set 1 and Set 2) and by dashed lines
for those with maximal stop mixing (Set 3 and Set 4), respectively. }
\end{figure}

Next, we discuss the one-loop corrections to the mass
relation~(\ref{eq:massrel}). Let us parameterize the deviation
from the tree-level relation by $\delta$, so that at the one-loop
order

\vspace*{-2mm}
\begin{eqnarray}
M_{H^\pm}^{} = \sqrt{M_{A}^2 + m_W^2} \left( 1 + \delta \right).
\label{eq:delta}
\end{eqnarray}

\vspace*{-2mm}
\noindent
We note that in our renormalization scheme (see Appendix~{\bf A}),
$M_A$ and $m_W$ are the input parameters, but $M_{H^+}$ is not.
The one-loop corrected mass of the charged Higgs boson
$M_{H^\pm}^{}$ can be obtained by solving
\begin{eqnarray}
 0 = {\rm Det} \left|
 \begin{array}{cc}
\Gamma^{(2)}_{G^+G^-}(p^2) & \Gamma^{(2)}_{G^+H^-}(p^2) \\
\Gamma^{(2)}_{H^+G^-}(p^2) & \Gamma^{(2)}_{H^+H^-}(p^2) \\
 \end{array} \right|,
\end{eqnarray}
where $\Gamma^{(2)}_{ij}(p^2)$ represent the renormalized
two-point functions in the basis of the renormalized Goldstone
boson ($G^\pm$) and charged Higgs boson ($H^\pm$) fields. Here,
The notation ``${\rm Det}$'' denotes taking the determinant of the
$2 \times 2$ matrix. One of the solution of the above equation is
$p^2=0$, which corresponds to the charged Nambu-Goldstone mode,
and another is $M_{H^\pm}^2$. At the one-loop level, the pole mass
of the charged-Higgs boson can be calculated from

\begin{eqnarray}
&&M_{H^\pm}^2 = M_A^2 + m_W^2 \nonumber \\
&&           + \Pi_{AA}(M_A^2) - \Pi_{H^+H^-}(M_A^2+m_W^2)
           + \Pi_{WW}(m_W^2), \label{eq:rmch}
\end{eqnarray}
where
$\Pi_{\phi\phi}(q^2)$ ($\phi=A,H^\pm$,and $W$) are
the self-energies.
For completeness, we list
the quark and squark contributions to the self-energies
of $A$ and $H^\pm$ in Appendix~{\bf E}.

When $A_{t,b}$ and $\mu$ are zero (i.e., no-mixing case),
the leading contribution (which is proportional to the forth power
of heavy quark mass) to $\delta$ is found to be
\vspace*{-0.2cm}
\begin{eqnarray}
\delta &\sim& \frac{N_c}{8\pi^2 v^2}
 \left(\frac{m_t^2 m_b^2}{M_{A}^2+m_W^2}\right)
\frac{1}{\sin^2\beta \cos^2\beta}
\left( 1 +\ln \frac{M^2}{m_t^2} \right).
\end{eqnarray}
This correction is substantial for $\tan\beta \simeq m_t/m_b$ and
$M^2 \gg m_t^2$. Applying Eq.~(\ref{eq:rmch}) with the complete
expression of $\Pi_{\phi\phi}(q^2)$, we found $\delta$ to be less
than 4.9\% for $2 < \tan\beta < 40$, $M < 2000$\,GeV and $M_A >
90$ GeV. Our result agrees well with Ref.~\cite{masses1}, in which
the approximate formula were presented for $M^2 \gg m_t^2$.

For the cases with a nonzero $A_{b,t}$ and $\mu$,
$\delta$ receives extra contributions, which are
proportional to $A_{t,b}^4/M^4$, $A_{t,b}^2 \mu^2/M^4$
and $\mu^4/M^4$\cite{masses2,masses1} originated from the squark
couplings [cf.
Eqs.~(\ref{coup1}), (\ref{coup2}), (\ref{coup3}) and (\ref{coup4})]
and squark masses.
For the Set 3 and Set 4 parameters listed in Table 1,
$\delta$ is less than 5.3\% and 3.6\% for $M_A > 90$ GeV, respectively.
In summary, as long as $|A_{t,b}|$ and $|\mu|$ are not too large
as compared to $M$, in a wide range of the parameter space
that is allowed by the
available experimental and theoretical constraints,
$\delta$ does not exceed 7-10\%.

Supported by the finding that
the one-loop electroweak corrections to the $W^\pm A H^\mp$ coupling
and to the mass relation $M^2_{H^\pm}=M_{A}^2+m_W^2$ are generally
smaller than the other theoretical errors
(such as the parton distribution function uncertainties)
and the expected experimental errors
(such as the mass resolution of Higgs boson decaying
into jets),
we anticipate that our conclusions
based upon a Born level analysis should also hold well
at the loop level.
Namely,
studying the process
 $ p {\bar p} \, , pp \to W^{\pm \ast} \to A H^\pm$
allows us to distinguish the MSSM from its alternatives by verifying
the mass relation~(\ref{eq:massrel}) and checking its
production rate.
If a signal is not found, studying this process
 can provide an upper bound
on the product of the decay branching ratios of $A$ and $H^\pm$ as
a function of the only one SUSY parameter -- $M_A$.

To detect the signal event, it is necessary to suppress its
potentially large backgrounds. For example,
the $\tau \nu b\bar b$ backgrounds can be largely reduced by having
a good $b$-tagging and
tau selection (by using the nature of $\tau$ polarization, which
differs between a parent $H^\pm$ and $W^\pm$ \cite{taunu}).
We expected that its observability is relatively easy at the LHC
and is a challenging task at the Tevatron for the signal event
rate is small.
Clearly, a detailed Monte Carlo analysis
is needed to calculate the significance of the signal event
at a collider.
This will be deferred to a future study.

\vspace*{1cm}

We thank K.~Hagiwara, G.-L.~Kane, P.~Nadolsky, 
Y.~Okada and T.~Tait for useful discussions.
This work was supported in part by the NSF grant PHY-9802564.

\section*{Appendices}
\subsection{Renormalization}

In this paper, we adopt the on-shell renormalization scheme
developed by Dabelstein\cite{dabelstein} to calculate the
one-loop electroweak corrections.
The standard model parameters are fixed by defining
$\alpha_{em}$, $m_W$ and $m_Z$, and the additional SUSY parameters
in the Higgs sector\footnote{
There are 7 parameters in the Higgs sector of the MSSM.
They are $g'$, $g$, $v_1$, $v_2$, $m_1$, $m_2$, and
$m_3$. Beyond the Born level, the wavefunction renormalization factors
$Z_{H_1}$ and $Z_{H_2}$ also need to be introduced to renormalize the
theory, where $H_1$ and $H_2$ denote the
two Higgs doublets in the model.
}
are fixed by the following renormalization
conditions:
(1) the tadpole contributions ($T_{H_1}=0$, $T_{H_2}=0$),
(2) the on-shell condition for the mass of $A$,
(3) the on-shell condition for the wavefunction of $A$,
(4) a renormalization condition on $\tan\beta$
   (which requires $\delta v_1/v_1=\delta v_2/v_2$), and
(5) a vanishing $A-Z$ mixing for an on-shell $A$.

\subsection{Calculation of $F^{(1)}(q^2)$}

The one-loop correction to
the renormalized form factor of the $W^\pm H^\mp A$ vertex,
apart from the effective weak gauge coupling ${\bar g}$,
can be written as

\begin{eqnarray}
F^{(1)}(q^2) &=& {\tilde Z}_{AA}^{1/2} {\tilde Z}_{H^+H^-}^{1/2} \left\{
1 + \delta F_{WHA}  \right.\nonumber\\
&&\left. + F_{WHA}^{\rm 1PI}(M_A^2,M_H^2,q^2) \right\}-1,
\label{eq:ffactor}
\end{eqnarray}
where ${\tilde Z}_{AA}$ and
${\tilde Z}_{H^+H^-}$ are the finite wavefunction
factors for the renormalizations of the
external Higgs bosons $A$ and $H^\pm$.
In our scheme,

\vspace*{-2mm}
\noindent
\begin{eqnarray}
  {\tilde Z}_{AA} &=& 1,\\
  {\tilde Z}_{H^+H^-} &=& 1 - \Pi_{H^+H^-}'({M_A^2}+m_{W}^2)+
                  \Pi_{AA}'(M_A^2) \, ,
\end{eqnarray}
where $\Pi_{AA}'(M_A^2)$ denotes taking the derivative of the
two point function $\Pi_{AA}(k^2)$ of the CP-odd scalar $A$
with respect to $k^2$ at $k^2=M_A^2$, etc.
The terms inside the curly bracket of Eq.~(\ref{eq:ffactor})
arise from the renormalized vertex function of $WHA$.
$F_{WHA}^{\rm 1PI}(p_A^2,p_H^2, q^2)$ represents the one-loop
contribution of the
one-particle-irreducible (1PI) diagrams with $p_A^2,p_H^2,q^2$
as the four-momentum square of the
incoming $A$, $H^\mp$ and $W^\pm$ particles, respectively.
$\delta F_{WHA}$ is the counterterm contribution resulting
from the field renormalization of $H^+$ and $A$:

\vspace*{-2mm}
\noindent
\begin{eqnarray}
 H^+A  \to   H^+A \left(1+{1 \over 2}\delta Z_{H^+}+
 {1 \over 2}\delta Z_{A}\right).
\end{eqnarray}
In terms of the independent counterterms fixed by the
renormalization scheme, the wavefunction counterterms
$\delta Z_{H^+}$ and $\delta Z_{A}$
can be written as
$(\sin^2 \beta) \delta Z_{H_1} + (\cos^2 \beta) \delta Z_{H_2}$
which is found to be equal to
$-\frac{1}{2}\Pi_{AA}'(M_A^2)$.
We note that in $\delta F_{WHA}$ the contributions
from the counterterms of the weak gauge coupling and the
wavefunction renormalization of the $W$-boson
are not included, because they should be combined with the $W$-boson
self energy contribution to derive the
 running weak gauge coupling $\bar g (q^2)$.
In our numerical calculation, we use
$$
{\bar g}^2 = 4{\sqrt{2}}m_W^2G_F \, .
$$
In summary, the one-loop electroweak
correction to $F^{(1)}(q^2)$ is found to be

\vspace*{-2mm}
\noindent
\begin{eqnarray}
  &&F^{(1)}(q^2)  \equiv F_{WHA}^{\rm 1PI}(M_A^2,M_{H^\pm}^2,q^2)\nonumber\\
  &&- \frac{1}{2} \Pi_{H^+H^-}'(M_A^2+m_W^2)
  - \frac{1}{2} \Pi_{AA}'(M_A^2) .
\label{eq:ff}
\end{eqnarray}
In the above equation, the top- and bottom-loop
contribution to $F_{WHA}^{\rm 1PI}$ is given by

\vspace*{-2mm}
\begin{eqnarray}
&&   F_{WHA}^{\rm 1PI (quark)}(q^2,p_A^2,p_H^2)   \nonumber\\
&&=\sum_{fff'=ttb,bbt} F_{WHA}^{fff'}(q^2,p_A^2,p_H^2),
\end{eqnarray}

\vspace*{-2mm}
\noindent
with

\vspace*{-2mm}
\begin{eqnarray}
&&  F_{WHA}^{fff'}(p_A^2,p_H^2,q^2)
    = + \frac{N_c}{16\pi^2} y_f^2
\left\{ p_A^2 C_{31}^{fff'} - p_H^2 C_{32}^{fff'} \right.\nonumber\\
&&
         + (2 p_A \cdot p_H - p_A^2) C_{33}^{fff'}
         - (2 p_A \cdot p_H - p_H^2) C_{34}^{fff'} \nonumber \\
&&         + (D + 2) (C_{35}^{fff'} - C_{36}^{fff'})
         + p_A^2 C_{21}^{fff'} - (2 p_A \cdot p_H
\nonumber \\
&&
+ p_H^2) C_{22}^{fff'}
         - 2 p_A^2 C_{23}^{fff'}  - (D-2) C_{24}^{fff'}
         - m_f^2 C_{11}^{fff'}  \nonumber\\
&&
   \left.  - \left( q^2 + m_f^2 \right) C_{12}^{fff'} \right\}
       - c_f \frac{1}{16\pi^2} y_f y_{f'} m_f m_{f'} C_{0}^{fff'}.
\label{eq:qform1}
\end{eqnarray}
where $c_f=+1$ and $-1$ for $fff'=ttb$ and $bbt$, respectively, and
$C_{ij}^{fff'}$ are defined in terms of
the Passarino-Veltman functions\cite{PV} with
\begin{eqnarray}
     C_{ij}^{fff'} =
     C_{ij} \left(p_A^2, p_H^2, (p_A+p_H)^2; m_f,m_f,m_{f'}\right).
\end{eqnarray}
The stop- and sbottom-loop contribution is given by
\vspace*{-2mm}
\begin{eqnarray}
&&  F_{WHA}^{\rm 1PI(squark)}
(p_A^2,p_H^2,q^2) = \frac{N_c}{16\pi^2} \sqrt{2} \sum_{i,j,k=1}^2
 \nonumber   \\
&&  \times  \left\{ U_{iL}^\ast D_{Lk}
    \left(i \lambda[\tilde{t}_j^\ast,\tilde{t}_i, A] \right)
    \lambda[\tilde{b}_k^\ast,\tilde{t}_j, H^-]
  \tilde{C}^{\tilde{t}_i\tilde{t}_j\tilde{b}_k} \right.\nonumber\\
&&  \left. -  U_{kL}^\ast D_{Li}
    \left(i \lambda[\tilde{b}_i^\ast,\tilde{b}_j, A] \right)
    \lambda[\tilde{b}_j^\ast,\tilde{t}_k, H^-]
  \tilde{C}^{\tilde{b}_i\tilde{b}_j\tilde{t}_k} \right\},
\label{eq:sqform1}
\end{eqnarray}

\vspace*{-2mm}
\noindent
where $U_{Ii}$, $D_{Ii}$ are the rotation
matrices for stops and sbottoms between
the weak eigenstate basis ($I=L,R$) and the mass eigenstate basis $(i=1,2)$,
respectively.
$\lambda[\tilde{f}_i^\ast,\tilde{f}_j', \phi_k]$ represents the
coefficient of the $\tilde{f}_i^\ast\tilde{f}_j' \phi_k$ interaction
in the MSSM Lagrangian, as listed in Appendix~{\bf C}, and
\vspace*{-2mm}
\noindent
\begin{eqnarray}
 \tilde{C}^{\tilde{f}_i\tilde{f}_j'\tilde{f}_k''}
= \left( C_{11} - C_{12} \right) \left(p_1^2,p_2^2,q^2;
     m_{\tilde{f}_i}, m_{\tilde{f}_j'}, m_{\tilde{f}_k''} \right).
\end{eqnarray}
The quark (top and bottom) and squark (stop and sbottom)
loop contributions to the self-energies
$\Pi_{AA}(q^2)$ and $\Pi_{H^+H^-}(q^2)$ can be found
in Appendix {\bf E}.

\subsection{Squark couplings with $H^\pm$ and $A$}

The mass eigenstates of the squarks
are related to their weak eigenstates by the rotation matrix
$O^{f\dagger}_{i\,I}$ with
$\tilde{f}_i = \sum_{I} O^{f\dagger}_{i\,I} \tilde{f}_I$,
where  $i=1,2$ and $I=L,R$; $O^{f}_{I\,i}=U_{I\,i}$ and $D_{I\,i}$
for $f=t$ and $b$, respectively. In terms of the mixing angles
$\theta_f$, we have
\begin{eqnarray}
 \left(\begin{array}{c} \tilde{f}_1 \\ \tilde{f}_2 \end{array}\right)
=\left( \begin{array}{cc} \cos \theta_f & \sin \theta_f \\
                    -\sin \theta_f & \cos \theta_f \\
  \end{array}\right)
 \left( \begin{array}{c} \tilde{f}_L \\ \tilde{f}_R \end{array} \right).
\end{eqnarray}
Here, we define the mixing angle $\theta_f$ so that
$\tilde{f}_1$ is lighter than $\tilde{f}_2$.

The coupling constants among the weak-eigenstate squarks
and the Higgs bosons are defined through the Lagrangian
\begin{eqnarray}
  {\cal L} = \cdot\cdot\cdot
 +  \lambda[\tilde{f}_I^\ast,\tilde{f}'_J,\phi,...]
                           \tilde{f}_I^\ast \tilde{f}'_J \phi,...
+ \cdot\cdot\cdot.
\end{eqnarray}
Hence, the coupling constants for the
mass-eigenstate squarks
are a linear combination of the couplings for
the weak-eigenstate squarks, and
\begin{eqnarray}
\lambda[\tilde{f}_i^\ast,\tilde{f}'_j,\phi,...]
 &=&
\lambda[
\tilde{f}_I^\ast O^{f}_{I\,i}, O^{f'\dagger}_{i\,I} \tilde{f}'_J,\phi,...]
\nonumber \\
  &=&O^{f}_{I\,i} O^{f'\dagger}_{i\,I}
\lambda[\tilde{f}_I^\ast,\tilde{f}'_J,\phi,...].
\end{eqnarray}
The relevant couplings
$\lambda[\tilde{f}_I^\ast,\tilde{f}'_J,\phi,...]$,
denoted as
$\lambda_{\tilde{f}_I^\ast \tilde{f}'_J \phi,...}$,
are listed below.

\vspace*{-2mm}
\begin{eqnarray}
&&\lambda_{\tilde{b}_L^\ast \tilde{t}_L H^-} =
\frac{-\sqrt{2}}{v}
(m_W^2 \sin 2 \beta - m_b^2 \tan\beta - m_t^2 \cot\beta) ,\\
&&\lambda_{\tilde{b}_L^\ast \tilde{t}_R H^-} =
\frac{\sqrt{2}m_t}{v}(A_t \cot\beta+\mu)   \label{coup1},\\
&&\lambda_{\tilde{b}_R^\ast \tilde{t}_L H^-} =
\frac{\sqrt{2}m_b}{v}(A_b \tan\beta+\mu)   \label{coup2},\\
&&\lambda_{\tilde{b}_R^\ast \tilde{t}_R H^-} =
\frac{2\sqrt{2} m_t m_b}{v \sin 2\beta},\\
&&\lambda_{\tilde{f}_L^\ast \tilde{f}_L A} =
\lambda_{\tilde{f}_R^\ast \tilde{f}_R A} =  0,
(\tilde{f}=\tilde{t},\tilde{b}),\\
&&\lambda_{\tilde{b}_L^\ast \tilde{b}_R A} =
\frac{i m_b}{v}(A_b \tan\beta+\mu) \label{coup3},\\
&&\lambda_{\tilde{t}_L^\ast \tilde{t}_R A} =
\frac{i m_t}{v}(A_t \cot\beta+\mu) \label{coup4},\\
&&\lambda_{\tilde{b}_L^\ast \tilde{b}_L A A} =
\frac{-m_b^2}{v^2} \tan^2\beta + \frac{g_Z^2}{4}(T_b-Q_b {s}_W^2)
\cos2\beta,\\
&&\lambda_{\tilde{b}_R^\ast \tilde{b}_R A A} =
\frac{-m_b^2}{v^2} \tan^2\beta + \frac{g_Z^2}{4}Q_b {s}_W^2
\cos2\beta  ,\\
&&\lambda_{\tilde{b}_L^\ast \tilde{b}_L A A} =
\frac{-m_t^2}{v^2} \cot^2\beta + \frac{g_Z^2}{4}(T_t-Q_t {s}_W^2)
\cos2\beta,\\
&&\lambda_{\tilde{b}_R^\ast \tilde{b}_R A A} =
\frac{-m_t^2}{v^2} \cot^2\beta + \frac{g_Z^2}{4}Q_t {s}_W^2
\cos2\beta  ,\\
&&\lambda_{\tilde{b}_L^\ast \tilde{b}_L H^+H^-} =
\frac{-2 m_t^2}{v^2} \tan^2\beta + \frac{g_Z^2}{2}(T_t-Q_t {s}_W^2)
\cos2\beta,\\
&&\lambda_{\tilde{b}_R^\ast \tilde{b}_R H^+H^-} =
\frac{-2 m_b^2}{v^2} \tan^2\beta + \frac{g_Z^2}{2}Q_b {s}_W^2
\cos2\beta  ,\\
&&\lambda_{\tilde{t}_L^\ast \tilde{t}_L H^+H^-} =
\frac{-2m_b^2}{v^2} \tan^2\beta + \frac{g_Z^2}{2}(T_b-Q_t {s}_W^2)
\cos2\beta,\\
&&\lambda_{\tilde{t}_R^\ast \tilde{t}_R H^+H^-} =
\frac{-2m_t^2}{v^2} \cot^2\beta + \frac{g_Z^2}{2}Q_t {s}_W^2
\cos2\beta,
\end{eqnarray}
where $T_t,T_b,Q_t$ and $Q_b$ are $\frac{1}{2},\frac{-1}{2},\frac{2}{3}$
and $\frac{-1}{3}$, respectively, and
\begin{eqnarray}
&&\lambda_{\tilde{t}_I^\ast \tilde{b}_J H^+} =
 \lambda_{\tilde{b}_I^\ast \tilde{t}_J H^-},\\
&&\lambda_{\tilde{f}_R^\ast \tilde{f}_L A} =
 -\lambda_{\tilde{f}_L^\ast \tilde{f}_R A},
\end{eqnarray}
for $I,J=L,R$ and $\tilde{f}=\tilde{t}, \tilde{b}$.

\subsection{Squark contributions to the $\rho$ parameter}

The squark one-loop contribution
to the $\rho$ parameter is given by

\vspace*{-2mm}
\begin{eqnarray}
  \Delta \rho = \rho -1 = - 4 \sqrt{2} G_F
       {\rm Re}[ \Delta \Pi^{11}_T(0) - \Delta \Pi^{33}_T(0) ],
\end{eqnarray}

\vspace*{-2mm}
\noindent
with\cite{pifunc}

\vspace*{-2mm}
\begin{eqnarray}
      \Delta \Pi^{11}_T(q^2) &=& \frac{N_c}{16\pi^2}
                       \sum_{f=t,b}  \sum_{i,j=1}^2
                       {T_{f_L}}^2 |O_{Li}^f|^2 |O_{Lj}^f|^2 \nonumber\\
 &&  \;\;\;\;\;\;\;\;\;\;\;\;\;\;\;\;\;\;\;\;\;\;\;\;\;
                     \times B(q^2;m_{\tilde{f}_i}^2,m_{\tilde{f}_j}^2),\\
      \Delta \Pi^{33}_T(q^2) &=& \frac{N_c}{32\pi^2}
                       \sum_{i,j=1}^2 |U_{Li}|^2 |D_{Lj}^f|^2
                       B(q^2;m_{\tilde{u}_i}^2,m_{\tilde{d}_j}^2),
\end{eqnarray}

\vspace*{-2mm}
\noindent
where $O_{Ii}^t=U_{Ii}$ and $O_{Ii}^b=D_{Ii}$;
$B(q^2;m_1^2,m_2^2) \equiv
A(m_1^2)+A(m_2^2)-4 B_{22}(q^2;m_1^2,m_2^2)$.
By using the expression
\begin{eqnarray}
B(0;m_1^2,m_2^2)=-\frac{1}{2}(m_1^2+m_2^2)
+\frac{m_1^2m_2^2}{m_1^2-m_2^2} \ln\frac{m_2^2}{m_1^2} \, ,
\end{eqnarray}
Eq.~(\ref{eq:rho}) is deduced under the assumption that
$M^2=M_Q^2 \simeq M_U^2 \simeq M_D^2 \gg m_t^2$ and
the stop mixing is large
($m_t|X_t|\simeq M^2$ and $m_b|X_b| \simeq 0$),
so that $m_{{\tilde t}_1} \sim {\cal O}(m_Z)$, which yields
$m_{{\tilde t}_2} \sim \sqrt{2} M$,
and $m_{{\tilde b}_1} \sim m_{{\tilde b}_2} \sim M$.

\subsection{Self energies}

The (top and bottom) quark-loop contributions to the self-energies
$\Pi_{AA}(q^2)$ and $\Pi_{H^+H^-}(q^2)$ are expressed
in terms of the Passarino-Veltman functions\cite{PV} by

\vspace*{-2mm}
\begin{eqnarray}
&&  \Pi_{AA}^{\rm quark} (q^2) = - \frac{N_c}{16\pi^2}\,
    \sum_{f=t,b}
   2  y_f^2 \left\{    q^2 \left( B_1(q^2,m_f,m_f)
\right.\right. \nonumber\\
&&   \left.                + B_{21}(q^2,m_f,m_f) \right)
                     + D \cdot B_{22}(q^2,m_f,m_f)\nonumber \\
&& \left.
                     + m_f^2  B_{0}(q^2,m_f,m_f)
                                \right\}, \label{eq:qform2} \\
&&  \Pi_{H^+H^-}^{\rm quark} (q^2) =
  - \frac{N_c}{16\pi^2}\,2\,(y_b^2 + y_t^2)
       \left\{       q^2  \left( B_1(q^2,m_b,m_t)
\right.\right.
\nonumber\\
&&  \left.\left.
                           +  B_{21}(q^2,m_b,m_t) \right)
                     + D \cdot B_{22}(q^2,m_b,m_t) \right\}, \nonumber\\
&&  - \frac{N_c}{16\pi^2}\,4\,y_b y_t m_b m_t
B_{0}(q^2,m_b,m_t).  \label{eq:qform3}
\end{eqnarray}

The stop- and sbottom-loop contributions are given by

\vspace*{-2mm}
\begin{eqnarray}
&&\Pi_{AA}^{\rm squark} (q^2) =
- \frac{N_c}{16\pi^2} \sum_{\tilde{f}=\tilde{t},\tilde{b}}\;
     \sum_{i,j=1}^2   \nonumber\\
&&\times \lambda[\tilde{f}_i^\ast,\tilde{f}_j, A]
       \lambda[\tilde{f}_j^\ast,\tilde{f}_i, A]
       B_0(q^2,m_{\tilde{f}_i},m_{\tilde{f}_j})\\
&& - \frac{N_c}{16\pi^2} 2 \sum_{\tilde{f}=\tilde{t},\tilde{b}}\;
     \sum_{i=1}^2 \lambda[\tilde{f}_i^\ast,\tilde{f}_i,A,A]
       A(m_{\tilde{f}_i}),  \label{eq:sqform2}\\
&&\Pi_{H^+H^-}^{\rm squark} (q^2) = - \frac{N_c}{16\pi^2}
       \sum_{i,j=1}^2  \nonumber\\
&&       \times \lambda[\tilde{t}_i^\ast,\tilde{b}_j, H^+]
       \lambda[\tilde{b}_j^\ast,\tilde{t}_i, H^-]
       B_0(q^2,m_{\tilde{t}_i},m_{\tilde{b}_j})\\
&& - \frac{N_c}{16\pi^2} \sum_{\tilde{f}=\tilde{t},\tilde{b}}\;
     \sum_{i=1}^2 \lambda[\tilde{f}_i^\ast,\tilde{f}_i,H^+,H^-]
       A(m_{\tilde{f}_i}).  \label{eq:sqform3}
\end{eqnarray}
The self-energy $\Pi_{WW}^{}(q^2)$ of the $W$ boson was already presented
in the literature. For example, the quark-loop contribution can be
found in Ref.~\cite{hhkm}, and the squark-loop contribution in
Ref.~\cite{pifunc}.

\end{document}